# A Review and Demonstration of The Essence of Chaos by Edward N. Lorenz

Dr. Robert M. Lurie

To understand a mathematical procedure, program it.   Conrad Wolfram  (Wolfram Blog 11/23/10 at TED)

The Essence of Chaos by Dr. Edward N. Lorenz is a marvelous exposition on chaos. In this book Dr. Lorenz, famous for his butterfly icon of chaos, gives a detailed description of a new and realistic model of chaos; the sliding of a board (a toboggan to those that live in snowy climes) and a sled down a "bumpy" hill (moguls to the snow aficionados). His text shows numerous figures which were calculated by him and this reviewer has formulated this model using Mathematica.

This is an update of an article published in Mathematica in Education and Research, Volume 11, Number 4, 2006, page 404. This journal is no longer published and the publisher/editor has given permission to the author for any use of the original article. In particular, modifications for compatibility with Version 8 of Mathematica have been made and many more aspects of the model are now included. Permission from the publisher of The Essence of Chaos (University of Washington Press) and Dr. Lorenz was also given to quote from the book and copy figures..

## 1. Introduction

**Dr. Edward N. Lorenz published** *The Essence of Chaos* **in 1993 based on a series of lectures at the University of Washington in 1990 sponsored by Jessie and John Danz. This book (ISBN 0-295-97514-8) is still the best exposition of chaos that this reviewer has seen and is written by one of the major contributors to the mathematics of chaos. This book includes a clear description of what is meant by chaos, and a number of famous examples. These are detailed sufficiently for a mathematical understanding. A major portion of the book introduces a new model of chaos that is developed in considerable detail. Of particular interest to students and users of** *Mathematica* **is the potential of programming these examples and thus being able to easily investigate the various models of chaos in any degree of detail desired. This review will delve into Dr. Lorenz's new model.**

**A brief description of the book which was on the flaps of the cover of the original hard bound book may be found at http://www.washington.edu/uwpress/search/books/LORESS.html.**
 **In the Preface, Lorenz states:**
**"My decision to convert the lectures into a book has been influenced by my conviction that chaos, along with its many associated concepts - strange attractors, basin boundaries, period-doubling bifurcations, and the like - can readily be under-stood and relished by readers who have no special mathematical or other scientific background ---.  I have placed the relevant mathematical equations and their derivations in an appendix, which need not be read for an understanding of the main text ---.  In any event, a good deal of less simple mathematics has gone into the production of the illustrations; most of them are end products of mathematical developments, subsequently converted into computer programs."**

**This paper develops and presents programs to be run with** *Mathematica* **that create the figures from the book. (Permission from both the publisher and the author has been generously granted to copy figures and quotations from the text.) The intent is that the readers will better understand the mathematics of the examples of chaos that Lorenz develops and they will be** allowed to experiment by varying the parameters and conditions. The formulation of the inputs stresses clarity rather than conciseness. The final preparation used *Mathematica* 8.0.4. Various sections of this paper clear the memory and re-define the variables and equations so that the separate sections can be run by themselves. It is obviously highly recommended that the book be available for the full understanding of this review ($16.95 list price).

## 2. Derivation of the Ski Slope

The main discussion in this book is a clear and unique presentation of a model of chaos: sleds or boards slide down a hill with moguls ("bumps of snow"). This model has a major advantage over many other models that have been used. Figure 4 (from the text) shows the mathematical model and Figure 5 (from the text) shows a photograph of a real ski slope. One can "feel" the effects with this real, physical example. In *Mathematica*, one can also develop animated programs that can give the reader a more real "experience", especially when compared to other models such as the logistic equation, Lorenz's own butterfly model, or the Restricted Three-Body Problem. In fact, Lorenz covers these and other examples of chaos in this very remarkable
book.

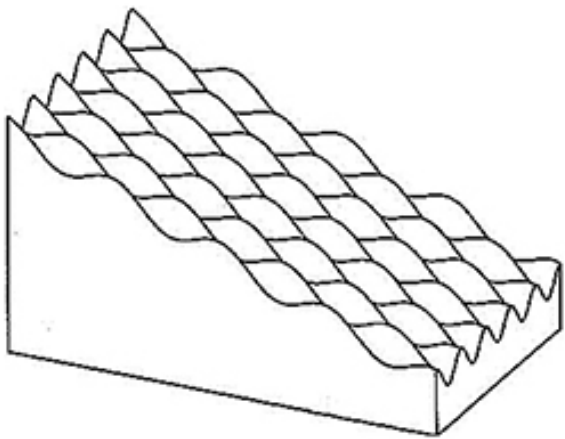

Figure 4. An oblique view of a section of the model ski slope.

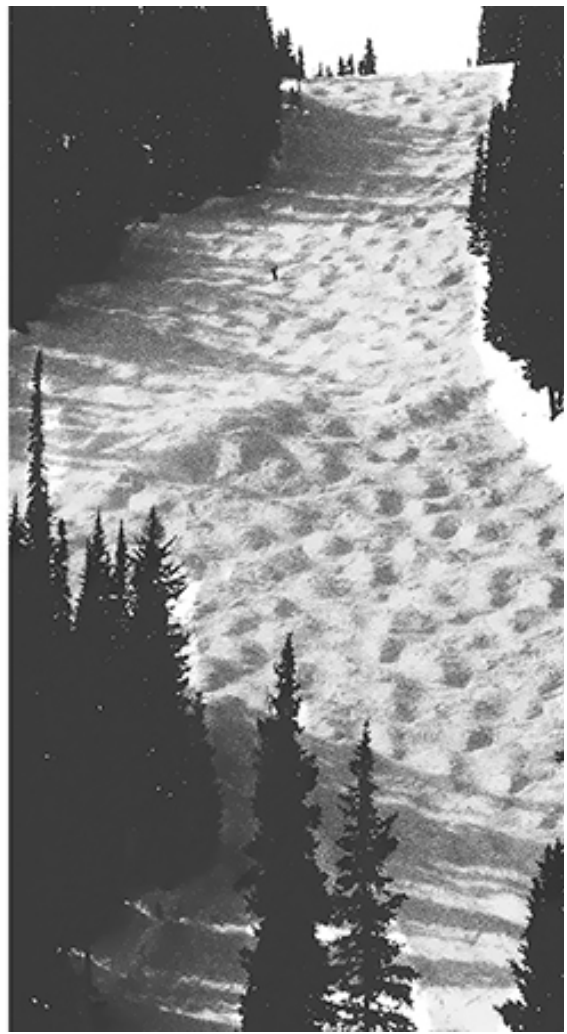

Figure 5 Moguls on a real-world ski slope.

The "relevant mathematical equations" are given in Appendix 2 starting on page 189. The vertical height of the ski slope is given by H; at the top of the hill H = 0

```
ClearAll["Global`*"]

H = -a * x - b * Cos[p * x] * Cos[q * y];
```

a is the slope of the hill. In most cases, Lorenz uses a value of 0.25; except in the Hamiltonian system, where a = 0, that is, no net slope for the hill. The last term of this equation provides the moguls. In his examples, $2\pi/p$ = 10.0 meters and $2\pi/p$ = 4.0 meters. In most of his examples, b = 0.5 meters, except for specified cases where b = 0.25 meters. In the discussion of bifurcation (not covered in this paper), b ranges from 0.0 to 0.6 meters. Note that h, the height of a mogul above a neighboring pit, is simply 2 b. "threeDmogul" is the name that this paper gives to the hill.

The height is multiplied by 1.02 so that the trajectories to be plotted are separated from the plot of the hill for maximum clarity..

```
a = 0.25; b = 0.5; p = 2*(Pi/10); q = 2*(Pi/4);
threeDmogulview1 = Plot3D[1.02*H, {x, 0, 30}, {y, -15, 15}, ViewPoint -> {2.6, -2.4, 2.}];

GraphicsRow[{threeDmogulview1}, ImageSize → 400]
```

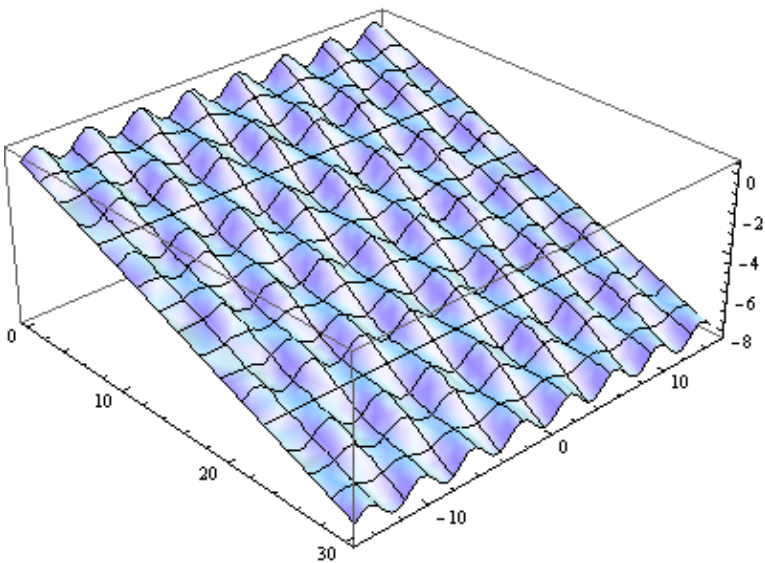

# 3. The Board Sliding on the Hill

Two types of "vehicles" are used to slide down this hill. The first is called a "board" which has a constant coefficient of friction. The other vehicle is called a "sled". The sled is controlled to have a constant downslope velocity, U. In this case, the coefficient of friction is chosen so that the differential of U, dU/dt is zero. For Hamiltonian systems the board is used with the friction set to zero and the slope of the hill also is set to zero.

As described on pages 189 and 190 of Lorenz, the basic equations are:

$\partial_t X = U$
$\partial_t Y = V$
$\partial_t Z = W$
$\partial_t U = -FH_x - cU$

$\partial_t\ V = -FH_y - cV$

$\partial_t\ W = -g + F - cW$

g is the acceleration of gravity (9.80665 Meters/seconds $^2$);

F is the vertical component of the force of the slope against the board or the sled;

c is the coefficient of friction;

subscripts denote partial differentiation.

Z=H(X,Y) =-a * x -b * Cos[p * x] * Cos[q * y]  (as seen above).

and it follows that   W=$H_x$U+$H_y$V, W being the velocity in the vertical direction.

p = Pi / 5;

q = Pi / 2;

a = 0.25;   this is the downward slope of the hill;  a=0 for Hamiltonian systems, that is, there is no net gravitational component

b= 0.5;      h=height of mogul = 2 * b

c = 0.5;  for board; for sled U=3.5 and c=-F0*Hx/U   and for Hamiltonian systems c=0

 On page 190 Lorenz says: "For the sled, U has been a prechosen constant, and c has been chosen to make dU/dt vanish ; thus c=-F0*Hx/U." That is, the friction is varied to keep the downward velocity constant; this could be done with some sort of brake.

**The basic conditions for this model is best sumarized on pages 30 and 32:**

**Since I have introduced the ski slope for the purpose of illustrating the fundamental properties of chaos, I shall choose values from a successful try. For convenience, let the slope face toward the south. Let its average vertical drop be 1 meter for every 4 meters southward. Imagine a huge checkerboard drawn on the slope, with "squares" 2 meters wide and 5 meters long, and let the centers of the moguls be located at the centers of the dark squares, as illustrated in Figure 6 (below), which also shows a possible path of a board down the slope. Let the damping time be 2 seconds. As detailed in Appendix 2, the formula selected for the topography of the slope will place pits in the light squares of the checkerboard, as if snow had been dug from them to build the moguls. Let each mogul rise 1 meter above the pits directly to its west and east. Like the pinball machine, the ski slope would have to be infinitely long to afford a perfect example of chaos.**

**We shall be encountering references to the four variables of the model so often that I shall give them concise names. These might as well consist of single letters, and they might as well be letters that have served as mathematical symbols in the differential equations, or in the computer programs for solving them. Let the southward and eastward distances of the board from some reference point, say the center of a particular pit, be called X and Y, respectively, and let the southward and eastward components of the velocity—the rates at which X and Y are currently increasing—be called U and V. Note that when we view the slope directly from the west side, as we nearly do in Figure 4 (above), or as we can do by giving Figure 6 a quarter turn counterclockwise, X and Y become conventional rectangular coordinates. Alternatively, but with some loss of mathematical convenience, we could have chosen the board's distance and direction from the reference point, and its speed and direction of motion, as the four variables. (Note: one usually expects to use time as the increment of "progress" or the independent variable; Lorenz here uses the distance down the hill. However in later examples (basin attractors --page 55 and Hamiltonian (no friction) - page 61) time is used as the independent variable.)**

**What I have been calling the centers of the moguls or the pits are actually the points where the slope extends farthest above or below a simple tilted plane. These are the centers of the checkerboard squares. The actual highest point in a dark square is about 1.5 meters north of the center of the mogul, while the lowest point in a light square is a like distance south of the center of the pit, and I shall call these points the high points and low points. They may be detected on the forward edge of the section of the slope in Figure 4.**

**To begin a computation, we can give the computer four numbers, which specify numerical values of X and Y, say in meters, and U and V, say in meters per second. In due time the computer will present us with more numbers, which specify the values of the same variables at any desired later times. As our first example, let X,Y, U, and V be 0.0, -0.5, 4.0, and 2.0, implying that the board starts half a meter due west of the center of a pit, and heads approximately south-southeastward. The board will then follow the sample path shown in Figure 6.**

**To see whether the board descends chaotically, let us turn to Figure 7, which shows the paths of seven boards, including the one appearing in Figure 6, as they travel 30 meters southward. All start from the same west-east starting line with the same speed and direction, but at points at 10-centimeter intervals, from 0.8 to 0.2 meters west of a pit. A tendency to be deflected away from the moguls is evident. The paths soon intersect, but the states are not alike, since now the boards are heading in different directions, and soon afterward the paths become farther apart than at first. By 10 meters from the starting line, the original 0.6-meter spread has more than doubled, and by 25 meters it has increased more than tenfold. Clearly the paths are**

**sensitively dependent on their starting points, and the motion is chaotic.**

**As already noted, an essential property of chaotic behavior is that nearby states will eventually diverge no matter how small the initial differences may be. In Figure 8 (not shown in this review), we let the seven boards travel 60 meters down the slope, starting from points spaced only a millimeter apart, from 0.503 to 0.497 meters west of the pit. At first the separation cannot be resolved by the picture, but by 30 meters it is easily detectable, and subsequently it grows as rapidly as in Figure 7. The ski slope has passed another critical test. The initial separation can be as small as we wish, provided that the slope is long enough. Lest the paths remind some of you of ski tracks, I should hasten to add that they do not follow routes that you, as knowledgeable skiers, would ordinarily choose. They might approximate paths that you would take if you fell and continued to slide.**

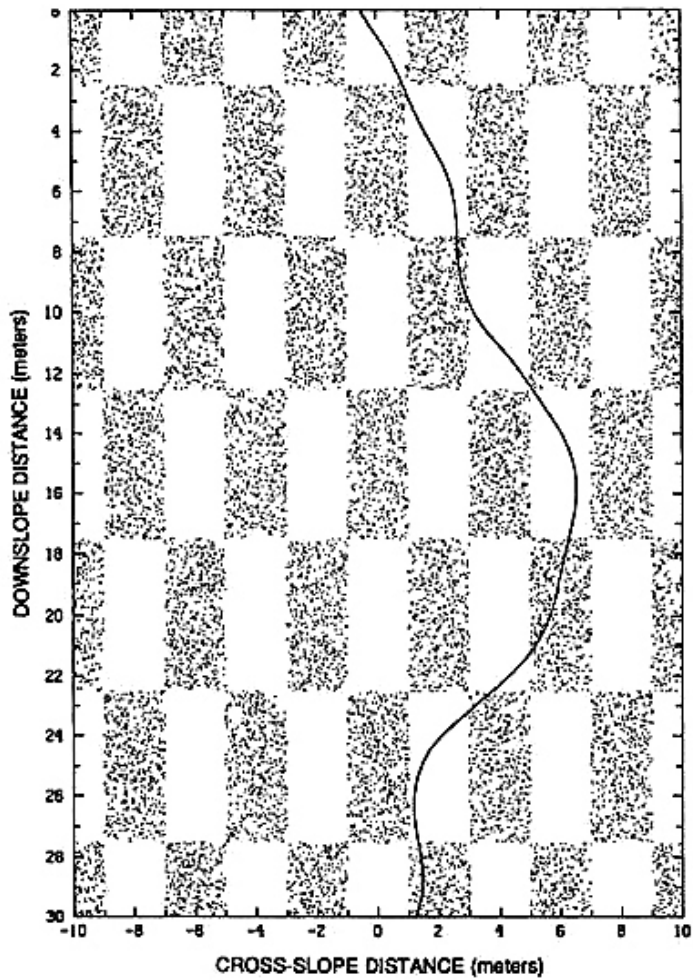

Figure 6. A top view of a section of the model ski slope, with the path of a single board sliding down it. The shaded rectangular areas of the slope project above a simple inclined plane, while the unshaded areas project below.

```
Hx = D[H, x];
Hy = D[H, y];
Hxx = D[H, x, x];
Hxy = D[H, x, y];
Hyy = D[H, y, y];
Z = H;
W = Hx*U + Hy*V;
Z = (-a)*x - b*Cos[p*x]*Cos[q*y];
F0 = (g + Hxx*U^2 + 2*Hxy*U*V + Hyy*V^2)/(1 + Hx^2 + Hy^2);
F1 = V/U;
H = -a * x - b * Cos[p * x] * Cos[q * y];
```

**F1 is delta y as dx moves  see p 190**

```
F2 = ((-F0)*Hy - c*V)/U;
```

**F2 is delta v as dx moves**

```
F4 = ((-F0)*Hx - c*U)/U;
```

**F4 is delta U as dx moves; but for sleds, U is constant**

**The value of various constants are as follows:**

```
p = Pi/5;
q = Pi/2;
a = 0.25;
b = 0.5;
c = 0.5;
```

**c= 0.5 for board;   for sled U = 3.5 and c = -F0*Hx/U; see page 190. c=0 for Hamiltonian cases.**

```
g = 9.80665;
reps = {y -> y[x], V -> V[x], U -> U[x]};
```

**reps = replacement; the function NDSolve will want to express y, V, U as functions of x and this replacement does that.
The following is an example of obtaining the path of the board going down the hill. The steps in the calculation are separated so that one can see the effects of each step.**

```
sol7 = NDSolve[{Derivative[1][y][x] == (F1 /. reps), Derivative[1][V][x] == (F2 /. reps),
    Derivative[1][U][x] == (F4 /. reps), y[0] == -0.5, V[0] == 2, U[0] == 4},
   {y, V, U}, {x, 0, 30}, MaxSteps -> 10 000];
height = (-a)*x - b*Cos[p*x]*Cos[q*y[x] /. sol7];
ynew7 = y[x] /. sol7;
lpts7 = Table[{x, ynew7, height}, {x, 0, 30, 0.001}];
flat7 = Flatten[lpts7];
part7 = Partition[flat7, 3];
scatterdata7a = ListPointPlot3D[part7, PlotStyle -> Directive[{PointSize[0.01], Hue[0.1]}]];
threeDmogul = Plot3D[1.02*H, {x, 0, 30}, {y, -15, 15}, PlotPoints -> 75,
   ViewPoint -> {3.015, 0., 1.536}, ImageSize -> {400, Automatic}];
Show[threeDmogul, scatterdata7a]
```

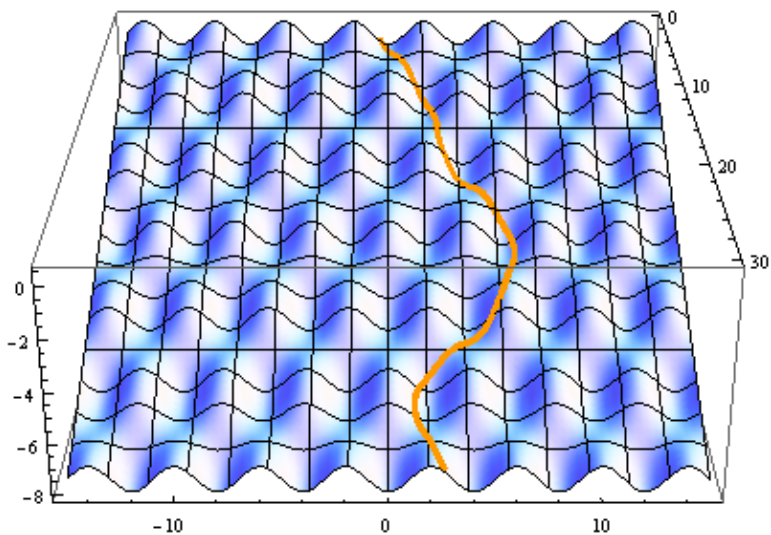

# 4. Multiple Boards on the Hill

To show the importance of starting conditions, the starting point for a series of boards are spaced at 10 centimeter intervals in the cross-hill direction, all other conditions remaining the same.

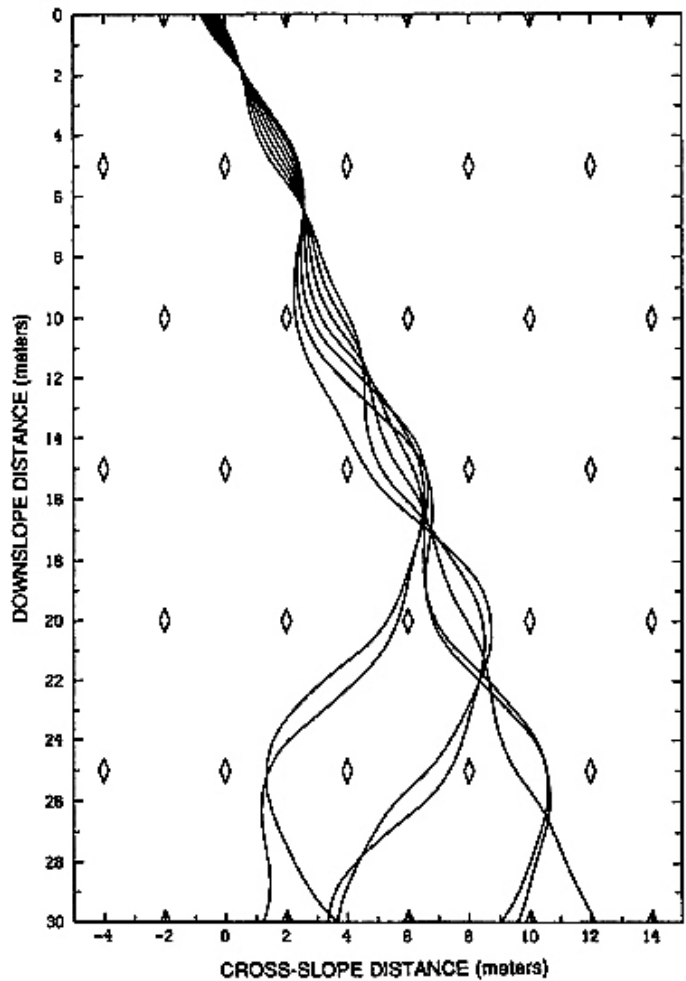

**Figure 7**. The paths of seven boards starting with identical velocities from points spaced at 10-centimeter intervals along a west-east line. The small diamonds indicate the locations of the centers of the moguls.

```
scatterdata[m_] := Module[{}, sol[m] = NDSolve[{Derivative[1][y][x] == (F1 /. reps),
    Derivative[1][V][x] == (F2 /. reps), Derivative[1][U][x] == (F4 /. reps),
    y[0] == -0.2 - 0.05*m, V[0] == 2, U[0] == 4}, {y, V, U}, {x, 0, 30}];
  ListPointPlot3D[Partition[Flatten[Table[{x, y[x] /. sol[m],
      (-a)*x - b*Cos[p*x]*Cos[q*y[x] /. sol[m]]}, {x, 0, 30, 0.01}]], 3],
   PlotStyle -> Directive[{Thickness[0.005], Hue[0.1*m]}]]];
```

```
Fig7 = Show[threeDmogul, scatterdata[2], scatterdata[4], scatterdata[5.85], scatterdata[8],
 scatterdata[10], scatterdata[0], scatterdata[12]]
```

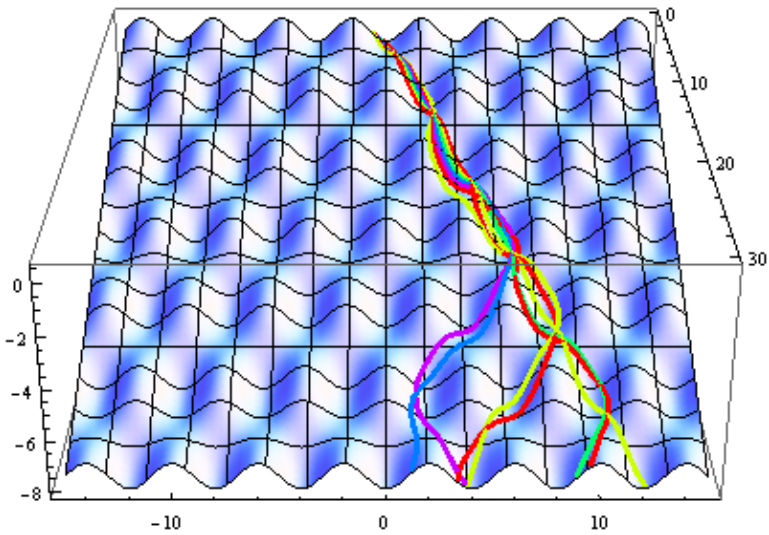

**We note that Figure 7 in the text is not quite duplicated unless the starting point for the board leaving at -.5 = (-.2-.05*6) is changed to -.49 = (-.2-.05*5.8). This may be due to differences in the programming and computers used by Lorenz and this author or perhaps an error in programming. (The same correction was needed when these figures were generated using Visual Basic.)**

**To show an animated version of two boards going down the hill that allows easy investigation of many of the variables and parameters the following Manipulate example combines many of the variables. Lorenz shows this in Figure 8 (not reproduced here) where the separations are 1 millimeter (versus 10 centimeters in the above example) and the hill length was increased to 60 meters.**

**(A simpler formulation has been shown at : http : // demonstrations.wolfram.com/ChaosWhileSleddingOnABumpySlope/)**

```mathematica
Manipulate[path3[b, d, If[m, 20, None], length, distanceDownSlope],
 {{b, 0.5, "mogul height"}, 0, 1, 0.05, Appearance -> "Labeled"},
 {{d, 0.2, "separation at start"}, 0, 1, 0.001, Appearance -> "Labeled"},
 {{length, 30, "length of hill"}, 30, 200, 10, Appearance -> "Labeled"},
 {{distanceDownSlope, 22}, 0, length, 0.5, Animator, AnimationRunning -> False,
  DefaultDuration -> 20}, {{m, True, "mesh"}, {True, False}},
 TrackedSymbols :> {b, d, m, length, distanceDownSlope},
 Initialization :> {path3[b_, d_, mesh_, length_, distanceDownSlope_] :=
   Module[
    {solution8, solution4, H, Hx, Hy, Hxx, Hxy, Hyy, W, U, F0, g, V, F1, F2, F4, q, a, c,
   reps, y, x, p, threeDmogul3}, H = (-a)*x - b*Cos[p*x]*Cos[q*y];
    Hx = D[H, x]; Hy = D[H, y];
   Hxx = D[H, x, x]; Hxy = D[H, x, y]; Hyy = D[H, y, y]; W = Hx*U + Hy*V;
   F0 = (g + Hxx*U^2 + 2*Hxy*U*V + Hyy*V^2)/(1 + Hx^2 + Hy^2); F1 = V/U;
   F2 = ((-F0)*Hy - c*V)/U; F4 = ((-F0)*Hx - c*U)/U; p = Pi/5; q = Pi/2; a = 0.25; c = 0.5;
   g = 9.80665; reps = {y -> y[x], V -> V[x], U -> U[x]};
   solution8 = Quiet[NDSolve[{Derivative[1][y][x] == (F1 /. reps), Derivative[1][V][x] ==
     (F2 /. reps), Derivative[1][U][x] == (F4 /. reps), y[0] == -0.4 - d, V[0] == 2,
    U[0] == 4}, {y, V, U}, {x, 0, length}]]; solution4 =

     Quiet[NDSolve[{Derivative[1][y][x] == (F1 /. reps), Derivative[1][V][x] == (F2 /. reps),
    Derivative[1][U][x] == (F4 /. reps), y[0] == -0.4, V[0] == 2, U[0] == 4}, {y, V, U},
    {x, 0, length}]]; threeDmogul3 = Plot3D[1.1*((-a)*x - b*Cos[p*x]*Cos[q*y]),
   {x, 0, length}, {y, -15, 25}, PlotPoints -> ControlActive[20, 50],
   Mesh -> ControlActive[None, mesh], MeshStyle -> Opacity[0.4],
   ViewPoint -> {4., -0.95, 2.45}, ImageSize -> {350, Automatic}, MaxRecursion -> 0];
   Show[threeDmogul3,
     ParametricPlot3D[Evaluate[{x, y[x], (-a)*x - b*Cos[p*x]*Cos[q*y[x]]} /.
    solution8], {x, 0, distanceDownSlope}, PlotRange -> {{0, length}, {-15, 15}, {-8, 0}},
   PlotStyle -> {Orange, Thickness[0.01]}], ParametricPlot3D[
   Evaluate[{x, y[x], (-a)*x - b*Cos[p*x]*Cos[q*y[x]]} /. solution4],
    {x, 0, distanceDownSlope}, PlotRange -> {{0, length}, {-15, 15}, {-8, 0}},
   PlotStyle -> {Black, Thickness[0.01]}], ImageSize -> {350, 300}, ImagePadding -> 20]]}]
```

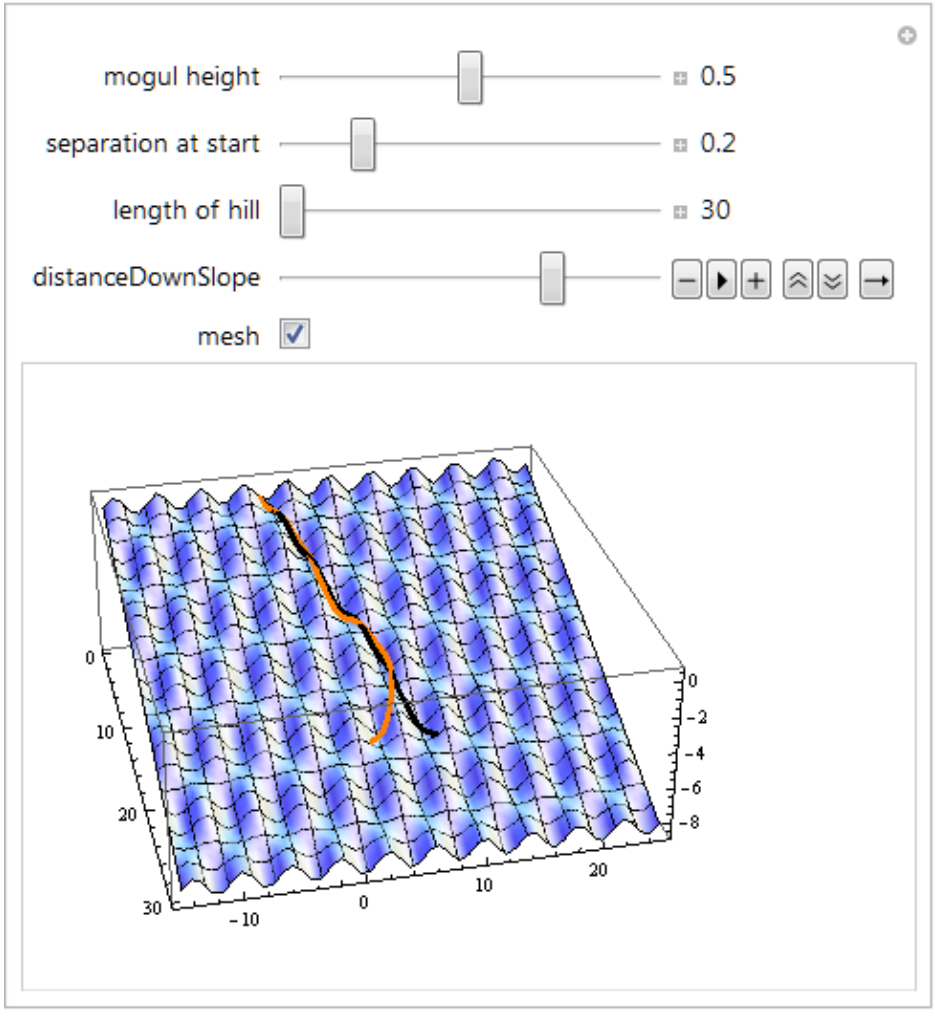

# 5. Strange Attractors

**The following are excerpts from the book, pages 37ff. It is perhaps the best statement of "strange attractors" that the reviewer has seen and worth careful reading in its complete form in the book.**

**An expression that has joined "chaos" in working its way into the scientific vocabulary, and that has aroused a fair amount of popular interest as well, is "strange attractor." Let us see what is meant by an attractor, and what one must be like to be considered strange.**

**The states of any system that do occur again and again, or are approximated again and again, more and more closely, therefore belong to a rather restricted set. This is the set of attractors.**

**When we perform a numerical experiment with a mathematical model, the same situation arises. We are free to choose any meaningful numbers as initial values of the variables, but after a while certain numbers or combinations of numbers may fail to appear. For the sled on the slope, with the value of U, the downslope speed, fixed at 3.5 meters per second, we can choose any initial value for V, the cross-slope speed, and any values for x and y within restricted ranges. Computations show that V will soon become restricted also, remaining between -5.0 and +5.0 meters per second.**

**Moreover, even the values of V that continue to occur will not do so in combination with certain values of the other variables. The sled will frequently slide almost directly over a mogul, and it will often move almost directly from the northwest or northeast, but whenever it is crossing a mogul it will be moving only from almost due north. To the computer, this means that, if x is close to 0.0 and y is close to -2.0 or +2.0, V will prove to be close to 0.0. Again, the states that do manage to occur, after**

the disappearance of any transient effects that may have been introduced by the choice of initial conditions, will form the set of attractors.

For the sled on the slope, it is convenient to use x, y, and V as coordinates in the three-dimensional phase space. The central region, containing the attractor, will then fit into a rectangular box—a box within which x extends only from -2.5 to +2.5 and y extends only from -2.0 to +2.0, since by definition x and y are limited to these ranges, while V extends only from -5.0 to +5.0, since larger cross-slope speeds do not occur, except transiently.

**... For our system, let us adopt the simple procedure of displaying cross sections of the box. The simplest of these sections are rectangles, parallel to one face or another. Mathematically, it is easiest to work with rectangles on which x is constant, and on which V is plotted against y. Horizontal and vertical distances from a central point of a rectangle can then equal values of eastward distance and eastward speed, respectively, just as they do in the phase space of the pendulum when the clock faces south. ... To determine the desired points, we may begin by taking any large collection of points on one of the rectangles. Each point will represent the initial state of one sled. Let us choose the rectangle on which x equals 2.5, so that the sleds all start from a west-east line that passes midway between a pit and the mogul directly to its south. The upper left panel in Figure 11, which, unlike the previous figures, is a picture of phase space instead of something tangible, contains five thousand points chosen at random. These are supposed to be a sampling of all points on the rectangle, representing the initial states of all sleds starting with any eastward speed between -5.0 and +5.0 meters per second, from any point on the starting line. The pattern has no recognizable form; it is true chaos in the non-technical sense of the word.**

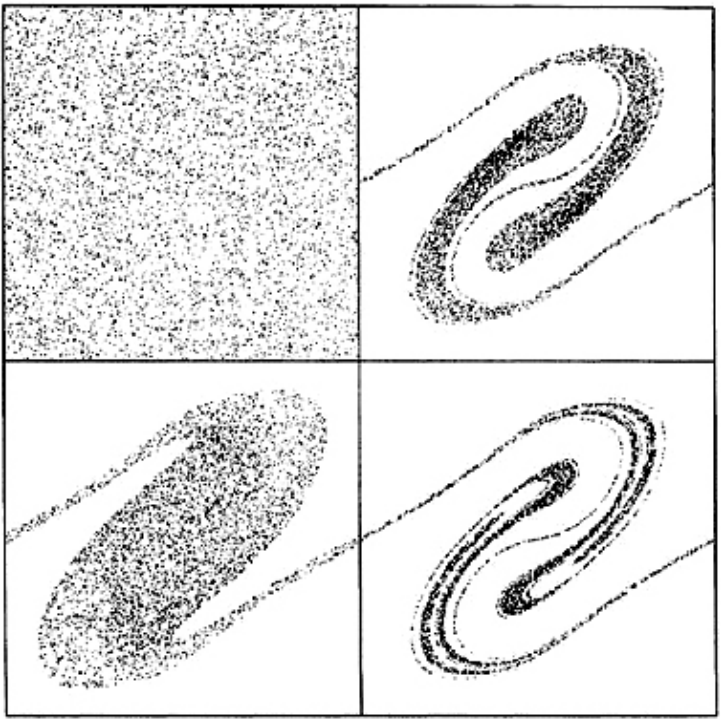

**Figure 11.** The upper left panel contains randomly chosen points representing the cross-slope positions and speeds of five-thousand sleds, all located on the same west-east line. The lower left and then the upper right and lower right panels represent the positions and speeds of the same sleds after they have traveled 5 and then 10 and 15 meters down the slope.

**We next allow each sled to descend 5 meters, so that x equals 2.5 again. In the lower left panel we have plotted the five thousand points representing the newly acquired cross-slope positions and velocities. The right-hand panels show what happens when the sleds have descended 10 and then 15 meters from the starting line. The points become attracted to regions that are more and more elongated and distorted.... The set to which these points will ultimately be**

**attracted, if we continue the process, will be the cross section of the attractor. We can more or less see what it will look like by extrapolating from the panels in Figure 11. The assemblage of points will become infinitely elongated, infinitesimally thin, and infinitely distorted...**

To duplicate Figure 11.

```
fnH[x_, y_] := -(a*x) - B*Cos[p*x]*Cos[q*y]
fnHx[x_, y_] := -a + B*p*Cos[q*y]*Sin[p*x]
fnHy[x_, y_] := B*q*Cos[p*x]*Sin[q*y]
fnHxx[x_, y_] := B*p^2*Cos[q*y]*Cos[p*x]
fnHxy[x_, y_] := -(B*p*q*Sin[q*y]*Sin[p*x])
fnHyy[x_, y_] := B*q^2*Cos[p*x]*Cos[q*y]
fnF0[v_, Hx_, Hy_, Hxx_, Hyy_, Hxy_, u_] :=
 (g + Hxx*u^2 + 2*Hxy*u*v + Hyy*v^2)/(1 + Hx^2 + Hy^2)
fnF1[v_, u_] = v/u;
fnF2[v_, c_, f0_, Hy_, u_] := (-(f0*Hy) - c*v)/u
p = Pi/5;
q = Pi/2;
a = 0.25;
B = 0.5;
u = 3.5;
c = -((f0*Hx)/u);
g = 9.80665;
H = fnH[x, y];
Hx = fnHx[x, y];
Hy = fnHy[x, y];
Hxx = fnHxx[x, y];
Hxy = fnHxy[x, y];
Hyy = fnHyy[x, y];
f0 = fnF0[v, Hx, Hy, Hxx, Hyy, Hxy, u];
f1 = fnF1[v, u];
f2 = fnF2[v, c, f0, Hy, u];
reps = {y -> y[x], v -> v[x], u -> 3.5};
sol11[yy_, vv_] := NDSolve[{Derivative[1][y][x] == (f1 /. reps),
   Derivative[1][v][x] == (f2 /. reps), y[-2.5] == yy, v[-2.5] == vv},
  {y, v}, {x, -2.5, 40}, MaxSteps -> 1500000]
data11[x_, yy_, vv_] := Transpose[{Mod[Thread[Partition[
         Flatten[Table[{x, y[x] /. sol11[yy, vv], v[x] /. sol11[yy, vv]}]], 3]][[2]], 4],
   Thread[Partition[Flatten[Table[{x, y[x] /. sol11[yy, vv], v[x] /. sol11[yy, vv]}]], 3]][[
     3]]}];

data11a[x_, yy_, vv_] := Transpose[{Mod[Thread[Partition[
         Flatten[Table[{x, y[x] /. sol11[yy, vv], v[x] /. sol11[yy, vv]}]], 3]][[2]] - 2, 4],
   Thread[Partition[Flatten[Table[{x, y[x] /. sol11[yy, vv], v[x] /. sol11[yy, vv]}]], 3]][[
     3]]}];

gridSize = 0.2;
```

**The gridSize determines how many points are plotted. The time for calculation increases substantially as the number of points, however, at gridSize =.1 Figure 11 is more closely duplicated.**

```
tabneg25 =
  Table[data11[xx, yy, vv], {xx, -2.5, -2.5}, {yy, -2, 2, gridSize}, {vv, -5, 5, gridSize}];
fig11neg25 = ListPlot[Partition[Flatten[tabneg25], 2], PlotStyle -> PointSize[0.005],
    PlotRange -> All, AspectRatio -> 1, Axes -> False];
```

```
tab25 =
  Table[data11a[xx, yy, vv], {xx, 2.5, 2.5}, {yy, -2, 2, gridSize}, {vv, -5, 5, gridSize}];
fig1125 = ListPlot[Partition[Flatten[tab25], 2], PlotStyle -> PointSize[0.005],
    PlotRange -> {-4, 4}, AspectRatio -> 1, Axes -> False];

tab75 = Table[data11[xx, yy - 2, vv],
    {xx, 7.5, 7.5}, {yy, -2, 2, gridSize}, {vv, -5, 5, gridSize}];
fig1175 = ListPlot[Partition[Flatten[tab75], 2], PlotStyle -> PointSize[0.005],
    PlotRange -> {-4, 4}, Axes -> False, AspectRatio -> 1];

tab125 =
  Table[data11a[xx, yy, vv], {xx, 12.5, 12.5}, {yy, -2, 2, gridSize}, {vv, -5, 5, gridSize}];
fig11125 = ListPlot[Partition[Flatten[tab125], 2], PlotStyle -> PointSize[0.005],
    PlotRange -> {-4, 4}, AspectRatio -> 1, Axes -> False];

GraphicsGrid[{{fig11neg25, fig1175}, {fig1125, fig11125}}]
```

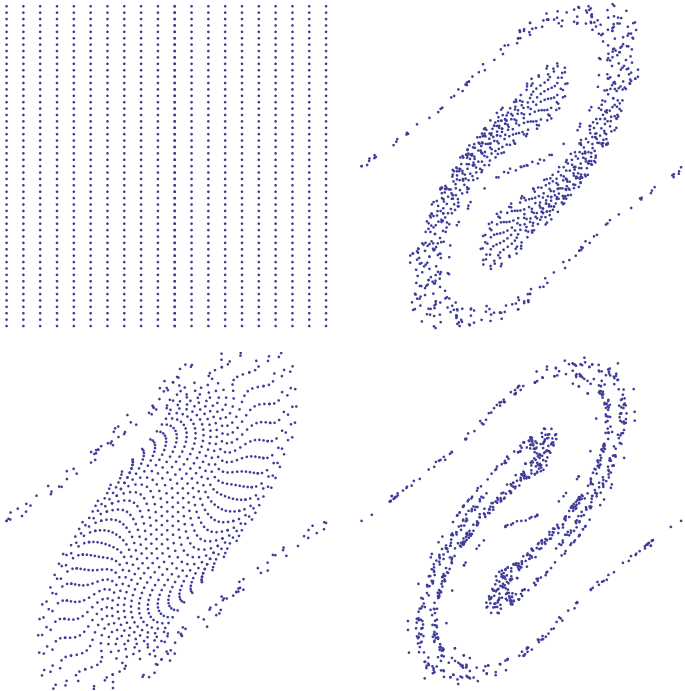

The following is a modified version of a demonstration: http://demonstrations.wolfram.com/SleddingOnABumpySlopeChaosAndStrangeAttractor/. One sled is coming down the hill and the Poincaré diagram is shown (across velocity on the y axis and the across position on the x axis. One can plot two different diagrams which have different downhill positions. Since the mogul pattern is 10 units (meters) in the downhill direction, the pattern repeats itself every 10 meters. The across hill position repeats itself every 4 meters and the plotting function contains a Mod 4 calculation to bring all similar positions together. If, for example, the slope of the hill is increased to 0.8 and the mogul height left at 0.5, then the diagrams show a nearly constant cross velocity-cross position which indicates a non-chaotic situation; this also is seen by the essentially repetitive path of the sled. With the mogul height at .15, the path is straight downhill.

```
Manipulate[p = Pi/5; q = Pi/2;
 H = (-hillslope)*x - mogulheight*Cos[p*x]*Cos[q*y]; length = 100;
 Hx = D[H, x]; Hy = D[H, y]; F0 = (g + D[H, x, x]*U^2 + 2*D[H, x, y]*U*V + D[H, y, y]*V^2)/
   (1 + Hx^2 + Hy^2); F1 = V/U; F2 = ((-F0)*Hy - ((-F0)*(Hx/U))*V)/U; U = 3.5; g = 9.80665;
 reps = {y -> y[x], V -> V[x], U -> 3.5};
 sol10 =
  Quiet[NDSolve[{Derivative[1][y][x] == (F1 /. reps), Derivative[1][V][x] == (F2 /. reps),
     y[0] == -0.6, V[0] == 2}, {y, V}, {x, 0, 5000}, MaxSteps -> 1500000]];
 Column[{Show[Plot3D[1.1*H, {x, 0, length}, {y, -15, 15}, PlotPoints -> 50,
     Mesh -> If[m, 15, None],
      ViewPoint -> {2, 0.2, 1}, PlotRange -> {{0, length}, {-15, 15},
      hillslope*(length/30)*{-35, 5}}, ImageSize -> {295, Automatic}],
     ListPointPlot3D[Partition[Flatten[Table[{x, y[x] /. sol10, (-hillslope)*x -
       mogulheight*Cos[p*x]*Cos[q*y[x] /. sol10]}, {x, 0, length, 0.01}]], 3],
     PlotRange -> {{0, length}, {-15, 15}, {-8*length, 0}},
     PlotStyle -> Directive[Black, PointSize[0.01]]]],

   ListPlot[Transpose[{Mod[Thread[Partition[Flatten[Table[{x, y[x] /. sol10, V[x] /. sol10},
        {x, startpt, 5000, 10}]], 3]][[2]], 4],
       Thread[Partition[Flatten[Table[{x, y[x] /. sol10, V[x] /. sol10}, {x, startpt, 5000,
        10}]], 3]][[3]]}], PlotStyle -> Directive[PointSize[0.01], Red],
     PlotLabel -> Row[{"downhill position = ", startpt}], PlotRange -> {{0, 4}, {-4, 4}},
     AxesOrigin -> {0, 0}, AspectRatio -> 0.6, AxesLabel -> {"across\nposition",
      "across\nvelocity"}, ImageSize -> {295, Automatic}],

   ListPlot[Transpose[{Mod[Thread[Partition[Flatten[Table[{x, y[x] /. sol10, V[x] /. sol10},
        {x, secstart, 5000, 10}]], 3]][[2]], 4],
       Thread[Partition[Flatten[Table[{x, y[x] /. sol10, V[x] /. sol10}, {x, secstart, 5000,
        10}]], 3]][[3]]}], PlotStyle -> Directive[PointSize[0.01], Red],
     PlotLabel -> Row[{"downhill position = ", secstart}], PlotRange -> {{0, 4}, {-4, 4}},
     AxesOrigin -> {0, 0}, AspectRatio -> 0.6, AxesLabel -> {"across\nposition",
      "across\nvelocity"}, ImageSize -> {250, Automatic}]}],
 {{startpt, 0, "downhill position of Poincaré diagram"},
  Range[0, 10], ControlType -> Setter},
 {{secstart, 5, "downhill position of second Poincaré diagram"}, 0, 10, 1,
  Appearance -> "Labeled"}, {{hillslope, 0.25, "slope of hill"}, 0.2, 0.8, 0.05,
  Appearance -> "Labeled"}, {{mogulheight, 0.5, "mogul height"}, 0, 0.8, 0.01,
  Appearance -> "Labeled"}, {{m, True, "mesh"}, {True, False}},
 TrackedSymbols :> {startpt, secstart, hillslope, mogulheight, m}, ContinuousAction -> False]
```

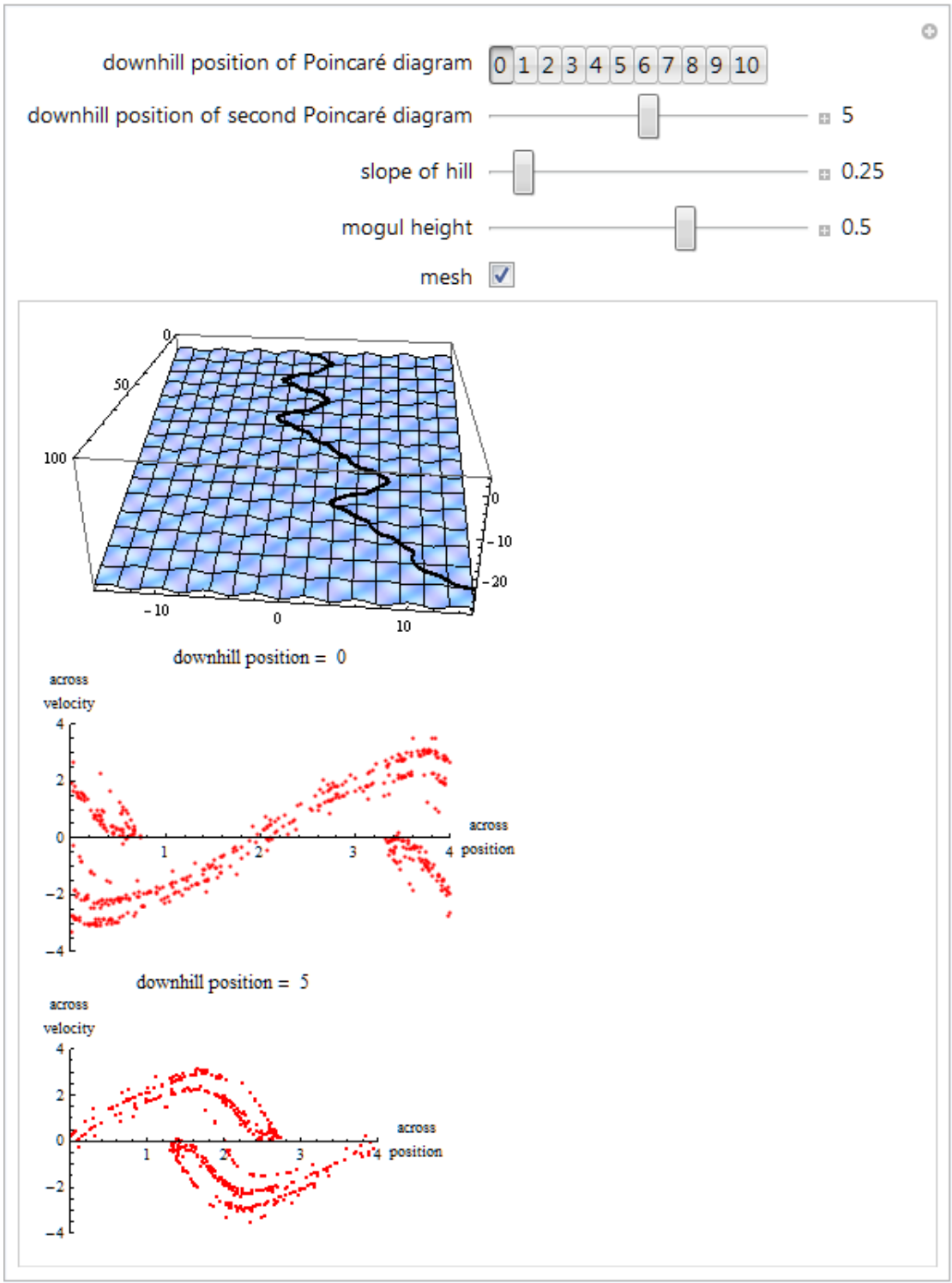

Figure 14 is a three - dimensional perspective view of the attreactor of the sled model, as pictured by nine parallel cross sections. "We can easily trace a number of features as they flow downward from one section to the next; a curve conecting similar features would represent a path taken by a sled. The complete attractor is seen to be composed of groups of nearly parallel surfaces, generally oriented more or less vertically; these are what appear as nearly parallel curves on each cross section. The continual stretching, compression, and twisting of the upper patterns to produce the lower ones is evident." (quoted from page 48).

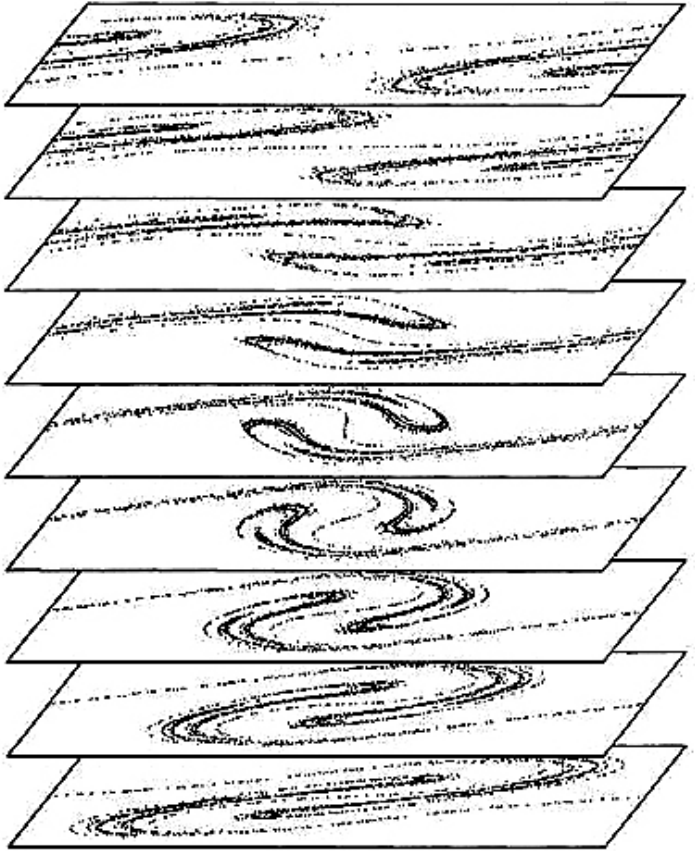

**Figure 14.** A three-dimensional perspective view of the attractor of the sled model, as pictured by nine parallel cross sections. The lowest section is the one appearing in Figures 12 and 13.

```
sol7 = NDSolve[
   {Derivative[1][y][x] == (F1 /. reps), Derivative[1][V][x] == (F2 /. reps), y[0] == -0.23,
   V[0] == 2}, {y, V}, {x, 0, 50 000}, MaxSteps -> 1 500 000];
poincare[sp_] := ListPlot[Transpose[
   {Mod[Thread[Partition[Flatten[Table[{x, y[x] /. sol7, V[x] /. sol7}, {x, sp, 50 000, 10}]],
     3]][[2]], 4], Thread[Partition[
       Flatten[Table[{x, y[x] /. sol7, V[x] /. sol7}, {x, sp, 50 000, 10}]],
     3]][[3]]]}, PlotStyle -> {PointSize[0.004], Hue[sp]}, PlotRange -> {All, All},
  AxesOrigin -> {0, 0},
 AspectRatio -> 0.6]
dataplot[sp_] := Transpose[{Mod[Thread[Partition[
       Flatten[Table[{x, y[x] /. sol7, V[x] /. sol7}, {x, sp, 50 000, 10}]], 3]][[2]],
   4], Thread[Partition[Flatten[Table[{x, y[x] /. sol7, V[x] /. sol7}, {x, sp, 50 000, 10}]],
     3]][[3]]]}
dataforplots = Table[dataplot[n], {n, 2.5, 7.5, 0.625}];
datafinal = MapIndexed[Append[#1, First[-#2]] & , dataforplots, {2}];
ListPointPlot3D[datafinal, PlotStyle -> PointSize[Tiny],
 BoxRatios -> {1, 1.5, 4}, ViewPoint -> {1.2, 10, 3},
 AxesLabel -> {"x", "y", "z"}, PlotRange -> All, ImageSize -> Large]
```

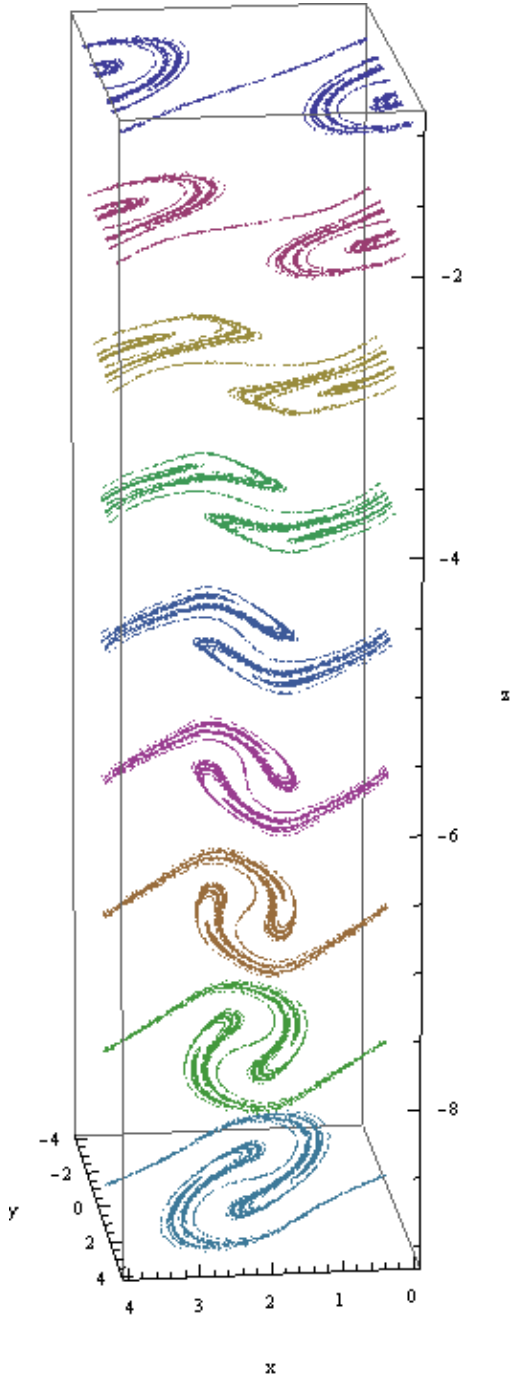

"In Figure 15, which, like the pictures of the attractors, is a diagram of a cross section of phase space, the points enclosed by the circle near the upper left corner represent the intiial positions and speeds of a collection of sleds; all of them are slightly east of a mogul and are moving a few degrees east of southward. The points enclosed by the curves that represent ellipses, and that lie successively farther from the circle, represent the positions and speeds of the same sleds after they have descended 1, 2, 3, 4, and then 5 meters. The continual stretching of one axis and compression of the other is apparent. The final ellipse", which is visibly distorted, looks almost like a segment of a curve." (quote from page 51)

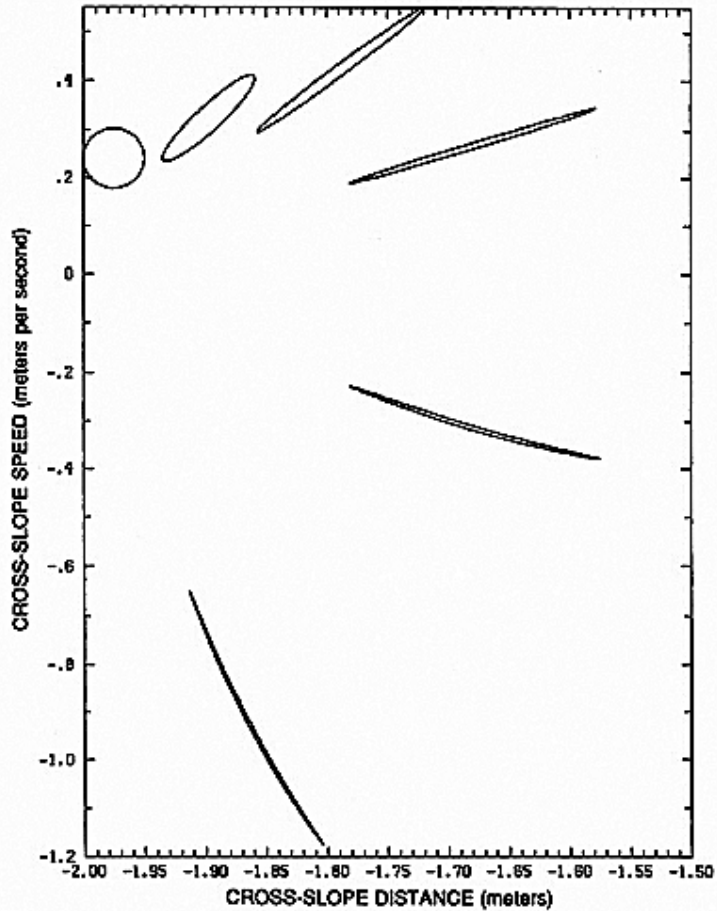

**Figure 15.** The points of the circle represent initial cross-slope positions and speeds of a collection of sleds. The elongated structures, in order of increasing distance from the circle, represent the positions and speeds of the same sleds at 1-meter intervals down the slope. Note that the figure covers only a small portion of the area covered by Figure 12.

```
fnH[x_, y_] := (-a)*x - B*Cos[p*x]*Cos[q*y]
fnHx[x_, y_] := -a + B*p*Cos[q*y]*Sin[p*x]
fnHy[x_, y_] := B*q*Cos[p*x]*Sin[q*y]
fnHxx[x_, y_] := B*p^2*Cos[q*y]*Cos[p*x]
fnHxy[x_, y_] := (-B)*p*q*Sin[q*y]*Sin[p*x]
fnHyy[x_, y_] := B*q^2*Cos[p*x]*Cos[q*y]
fnF0[v_, Hx_, Hy_, Hxx_, Hyy_, Hxy_, u_] :=
  (g + Hxx*u^2 + 2*Hxy*u*v + Hyy*v^2)/(1 + Hx^2 + Hy^2)
fnF1[v_, u_] = v/u;
fnF2[v_, c_, f0_, Hy_, u_] := ((-f0)*Hy - c*v)/u
p = Pi/5;
q = Pi/2;
a = 0.25;
B = 0.5;
u = 3.5;
c = (-f0)*(Hx/u);
g = 9.80665;
H = fnH[x, y];
Hx = fnHx[x, y];
Hy = fnHy[x, y];
Hxx = fnHxx[x, y];
Hxy = fnHxy[x, y];
Hyy = fnHyy[x, y];
f0 = fnF0[v, Hx, Hy, Hxx, Hyy, Hxy, u];
f1 = fnF1[v, u];
f2 = fnF2[v, c, f0, Hy, u];
reps = {y -> y[x], v -> v[x], u -> 3.5};
```

**Multiple sleds are sent down the hill starting in a circular array**.

```
data6[m_, t_] := Module[{}, sol[m] = NDSolve[{Derivative[1][y][x] == (f1 /. reps),
    Derivative[1][v][x] == (f2 /. reps), y[0] == -2 + 0.005*m,
    v[0] == 0.24 + ((1 - ((-2 + 0.005*m) + 1.975)^2/0.025^2)*0.06^2)^0.5},
     {y, v}, {x, 0, 30},
    MaxSteps -> 200 000]; Partition[Flatten[
     Table[{y[x] /. sol[m], v[x] /. sol[m]}, {x, 0, t, 1}]], 2]]
data7[m_, t_] := Module[{}, sol[m] = NDSolve[{Derivative[1][y][x] == (f1 /. reps),
    Derivative[1][v][x] == (f2 /. reps), y[0] == -2 + 0.005*m,
    v[0] == 0.24 - ((1 - ((-2 + 0.005*m) + 1.975)^2/0.025^2)*0.06^2)^0.5},
     {y, v}, {x, 0, 30},
    MaxSteps -> 200 000]; Partition[Flatten[
     Table[{y[x] /. sol[m], v[x] /. sol[m]}, {x, 0, t, 1}]], 2]]
```

```
datacurve15more = ListPlot[
   {data6[0, 5], data6[0.5, 5], data6[1, 5], data6[1.5, 5], data6[2, 5], data6[2.5, 5],
  data6[3, 5], data6[3.5, 5], data6[4, 5], data6[4.5, 5],
    data6[5, 5], data6[5.5, 5], data6[6, 5],
  data6[6.5, 5], data6[7, 5], data6[7.5, 5], data6[8, 5],
    data6[8.5, 5], data6[9, 5], data6[9.5, 5],
  data6[9.99, 5]}, AspectRatio -> 1.6];
datacurve25more = ListPlot[
   {data7[0, 5], data7[0.5, 5], data7[1, 5], data7[1.5, 5], data7[2, 5], data7[2.5, 5],
  data7[3, 5], data7[3.5, 5], data7[4, 5], data7[4.5, 5],
    data7[5, 5], data7[5.5, 5], data7[6, 5],
  data7[6.5, 5], data7[7, 5], data7[7.5, 5], data7[8, 5],
    data7[8.5, 5], data7[9, 5], data7[9.5, 5],
  data7[9.99, 5]}, AspectRatio -> 1.6];
```

**And this is Figure 15** :

```
Show[datacurve15more, datacurve25more]
```

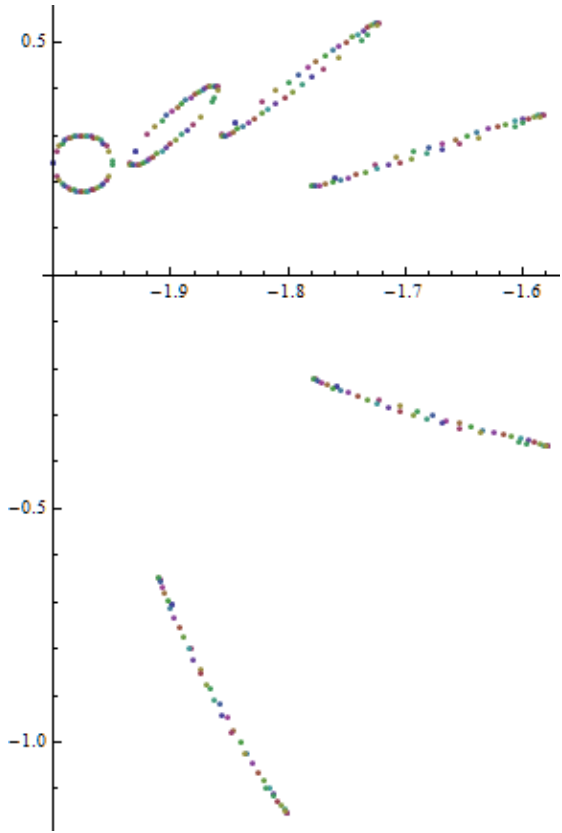

**By adding "Joined->True", the points for each sled are connected so that the paths are shown.**

```
datacurve15moreaa =
 ListPlot[{data6[0, 5], data6[0.5, 5], data6[1, 5], data6[1.5, 5], data6[2, 5],
  data6[2.5, 5], data6[3, 5], data6[3.5, 5],
    data6[4, 5], data6[4.5, 5], data6[5, 5], data6[5.5, 5],
  data6[6, 5], data6[6.5, 5], data6[7, 5], data6[7.5, 5],
    data6[8, 5], data6[8.5, 5], data6[9, 5],
  data6[9.5, 5], data6[9.99, 5]}, AspectRatio -> 1.6, Joined -> True, PlotMarkers -> Automatic]
```

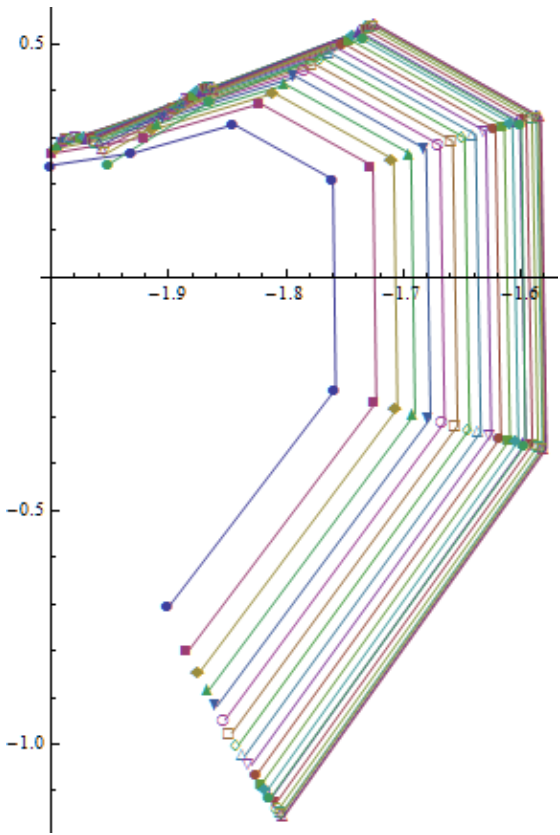

**For Figure 17, page 56 :Unlike the sled model, the board model has a second attractor. A board that starts moving very slowly near the low point of a pit, or one that is given an initial upward push and just manages to enter a pit, can become trapped, in which case it will eventually come to rest at the low point. The state of stable equilibrium that it attains is an attractor, represented in four-dimensional phase space by a single point.The model therefore does not always behave chaotically.
In Figure 17 we see portions of a few of the dark and light checkerboard squares (i.e. moguls) that cover the ski slope. We also see a specially constructed closed curve. A board starting from rest at any point enclosed by the curve will eventually come to rest again at the low point, indicated by the central dot. Boards starting from rest at points outside the curve, but still within the realm of Figure 17, will travel down the slope. On a more extensive picture of the slope, a similar closed curve would surround each pit.**

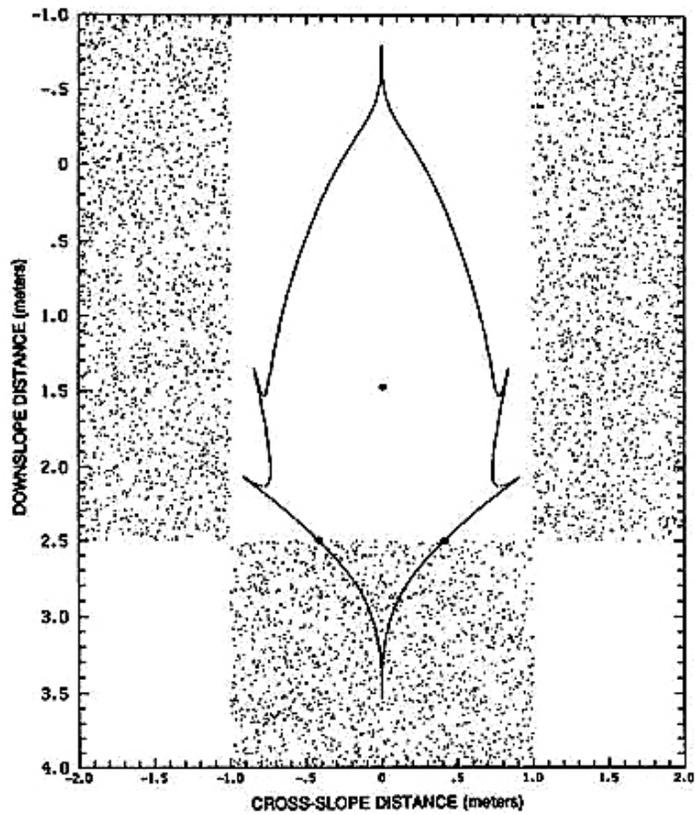

Figure 17. A section of the ski slope. The jagged curve separates the points on the slope from which a board starting from rest will become trapped in a pit from those from which it will continue down the slope. The central dot indicates the lowest point in the pit, and the other two dots indicate the saddle points.

```
Clear[x, y, u, v]
```

```
fnH[x_, y_] := (-a)*x - B*Cos[p*x]*Cos[q*y]
fnHx[x_, y_] := -a + B*p*Cos[q*y]*Sin[p*x]
fnHy[x_, y_] := B*q*Cos[p*x]*Sin[q*y]
fnHxx[x_, y_] := B*p^2*Cos[q*y]*Cos[p*x]
fnHxy[x_, y_] := (-B)*p*q*Sin[q*y]*Sin[p*x]
fnHyy[x_, y_] := B*q^2*Cos[p*x]*Cos[q*y]
fnF0[v_, Hx_, Hy_, Hxx_, Hyy_, Hxy_, u_] :=
  (g + Hxx*u^2 + 2*Hxy*u*v + Hyy*v^2)/(1 + Hx^2 + Hy^2)
fnF1[v_] := v
fnF3[u_] := u
fnF2[v_, c_, f0_, Hy_] := (-f0)*Hy - c*v
fnF4[u_, c_, f0_, Hx_] := (-f0)*Hx - c*u
p = Pi/5;
q = Pi/2;
a = 0.25;
B = 0.5;
c = 0.5;
g = 9.80665;
H = fnH[x, y];
Hx = fnHx[x, y];
Hy = fnHy[x, y];
Hxx = fnHxx[x, y];
Hxy = fnHxy[x, y];
Hyy = fnHyy[x, y];
f0 = fnF0[v, Hx, Hy, Hxx, Hyy, Hxy, u];
f1 = fnF1[v];
f3 = fnF3[u];
f2 = fnF2[v, c, f0, Hy];
f4 = fnF4[u, c, f0, Hx];
reps = {y -> y[t], x -> x[t], u -> u[t], v -> v[t]};
```

**Note that the integration is based on the progression of time rather than the progress in the x direction.**

```
figure17data[yyy_, xxx_] := Module[{},
  sol = NDSolve[{Derivative[1][y][t] == (f1 /. reps), Derivative[1][x][t] == (f3 /. reps),
    Derivative[1][u][t] == (f4 /. reps), Derivative[1][v][t] == (f2 /. reps), y[0] == yyy,
    u[0] == 0, v[0] == 0, x[0] == xxx}, {x[t], y[t], u[t], v[t]}, {t, 0, 20},
   MaxSteps -> 100 000]; Partition[Flatten[Table[{xxx, yyy, x[t] /. sol, y[t] /. sol,
     (-a)*x[t] - B*Cos[p*x[t]]*Cos[q*y[t]] /. sol}, {t, 20, 20}]], 5]];
```

**The above is only half of the figure since it is symmetrical.**
**The following may take 2 to 3 minutes run time depending on the computer.**

```
tablefig17 =
  Partition[Flatten[Table[figure17data[y3, x3], {y3, -1, 0, 0.02}, {x3, -1, 4, 0.02}]],
   5];
selectdata = Select[tablefig17, #1[[3]] < 4 &];
```

**The above equation selects those points with x less than 4; that is, those that did not leave the pit.**

```
selectdata2 = Table[{selectdata[[n]][[1]], selectdata[[n]][[2]], (-a)*selectdata[[n]][[1]] -
   B*Cos[p*selectdata[[n]][[1]]]*Cos[q*selectdata[[n]][[2]]]}, {n, 1, Length[selectdata]}];
negsselectdata2 = Table[{selectdata[[n]][[1]], -selectdata[[n]][[2]],
   (-a)*selectdata[[n]][[1]] - B*Cos[p*selectdata[[n]][[1]]]*Cos[q*selectdata[[n]][[2]]]},
  {n, 1, Length[selectdata]}];
```

**The above equation is the other half of the plot since it is symetrical.**

```
threeDmogul10 = Plot3D[1.06*((-a)*x - B*Cos[p*x]*Cos[q*y]),
   {x, -4, 4}, {y, -2, 2}, MeshStyle -> Opacity[0.5],
 ViewPoint -> {4., -0.95, 4.45}];

figure17plot = ListPointPlot3D[{selectdata2, negsselectdata2},
  PlotRange -> {{-1, 4}, {-2, 2}, {-1.5, 1}}, PlotStyle -> AbsolutePointSize[0.02],
  ViewPoint -> {4, -0.95, 5}];
Show[ figure17plot, threeDmogul10, AxesLabel -> {"x", "y", "z"}]
```

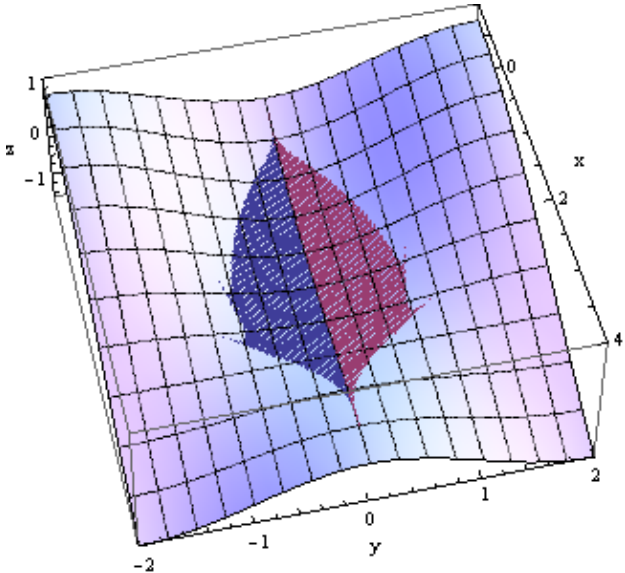

# 6. Chaos of Another Species (Hamiltonian)

**From page 60 ff :**
Models of familiar real physical systems that simply disregard all dissipative processes and all energy sources are generally Hamiltonian. Probably the most familiar real - world, or real - universe, Hamiltonian system consists of the sun with its planets orbiting about it.
Hamiltonian systems may be chaotic; note that the qualitative reasoning indicating that a pinball machine should behave chaotically does not invoke any dissipation. Chaotic systems therefore do not always possess strange attractors, although most of the generally encountered dissipative chaotic systems do have them.
To discover why this should be so, let us convert the board on the ski slope into a Hamiltonian system.We can do this, mathematically, by removing the friction, and also removing the general southward drop in elevation, which plays a similar role to the clockwork that drives the pendulum.The moguls and pits will then project from a horizontal surface.

The energy of a board consists of kinetic energy, represented by its speed, and potential energy, represented by its elevation above the bottom of a pit, and it is the sum of these two forms of energy that remains fixed as time advances.The total energy is therefore a constant of the model ....Let us consider in detail the possible behavior of a collection of boards, all having the same total energy .....For definiteness let us examine an intermediate in which the energy is nine - tenths of the minimum amount needed to reach the top of a mogul.....Figure 19 shows what can happen; here the starting points are 1 centimeter apart, and appear in the figure to be all the same. Before 50 meters the paths have visibly diverged, confirming the presence of chaos, but thereafter they proceed so erratically that when they are displayed together it is not easy to see which ending follows from which beginning.Two of the boards even turn around and

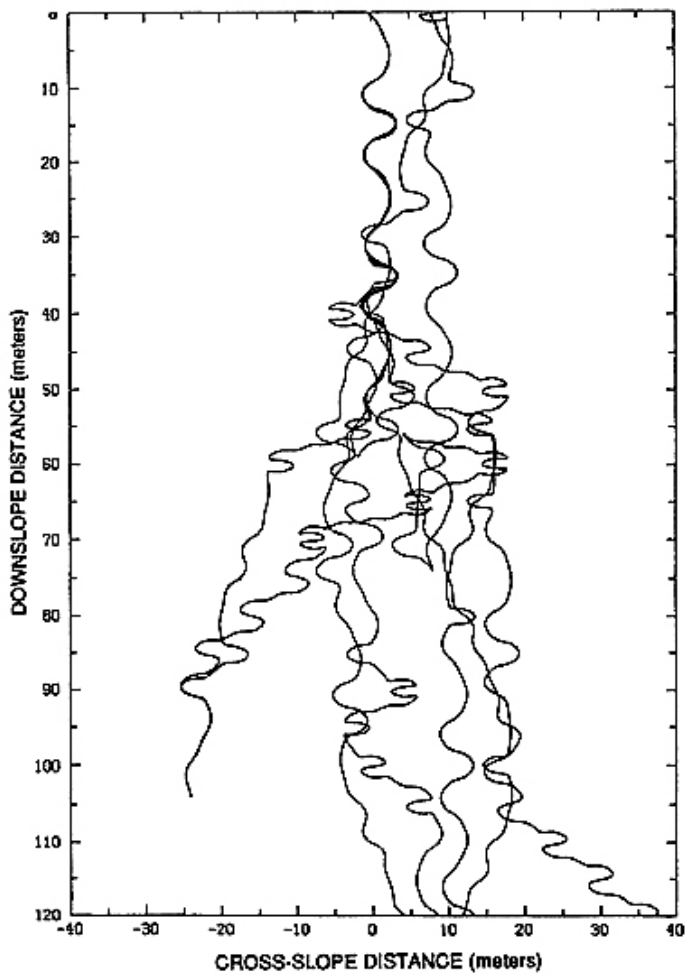

**Figure 19.** The paths of seven boards starting with identical total energies and identical velocities from points spaced at 1-centimeter intervals, on the Hamiltonian ski "slope" with no friction and no continual southward drop.

**slide back across the starting line—something that could not possibly happen in the dissipative model....There are various ways to display the long - term properties of these and other paths, but it is particularly simple to do just as we did with the dissipative model. We choose an initial point, and then plot the cross - slope speed against the cross - slope position, this time whenever the board crosses a west - east line through a pit and a mogul, i.e., whenever x equals 0. For the dissipative system the procedure produced a cross section of a strange attractor; we have already seen that for any Hamiltonian system it must do something else, since there will not be an attractor. Figure 20 shows what happens when the initial point is the start of one of the paths in Figure 19. Not surprisingly, the points appear to fill an area, sometimes called a chaotic sea, instead of lying on separate curves with gaps between them.What may surprise us is the four prominent holes.It seems rather unlikely that such large areas would be missed if they were to be eventually occupied.**

**If it is true that the sequences of points will never enter a hole, it must be equally true that other sequences beginning in a hole will never enter the sea.....Let us therefore repeat our procedure a number of times, choosing in each instance an initial state in one of the holes.We obtain the composite picture in Figure 21, containing four patches that would fit into the holes in Figure 20. Each new sequence produces a closed loop, or else a chain of small loops surrounding a larger loop.**

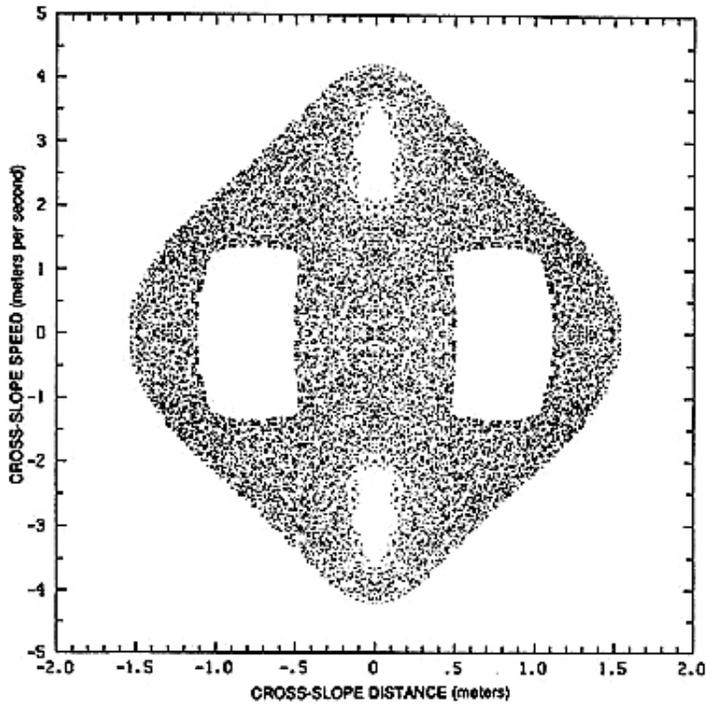

Figure 20. A Poincaré section of a chaotic sea produced by the Hamiltonian ski-slope model, when the total energy is nine-tenths of that needed to reach the top of a mogul.

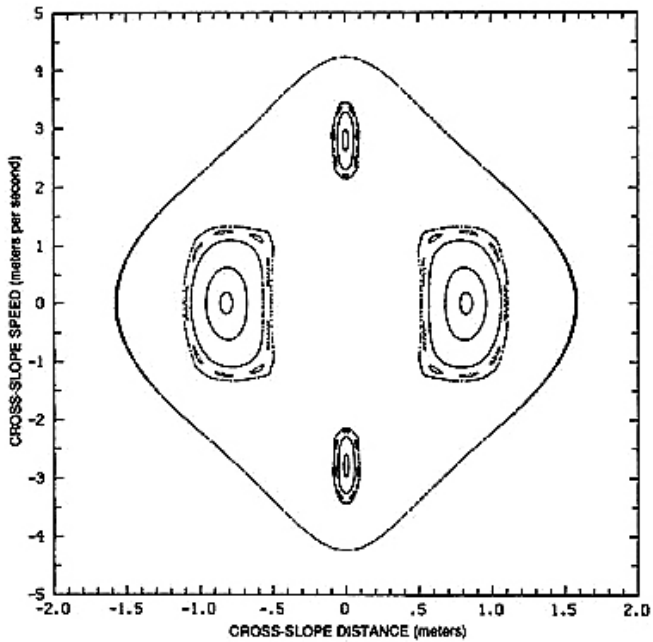

Figure 21. Some of the periodic islands and a periodic shoreline of the chaotic sea of Figure 20.

**The following programs generate Figures 19 to 21 as described above. Note that the equations are based on time being varied rather than having the x (downhill) direction being the independent variable.**

```
Clear["Global`*"]
```

```
H = (-a)*x - B*Cos[p*x]*Cos[q*y];
Hx = D[H, x];
Hy = D[H, y];
Hxx = D[H, x, x];
Hxy = D[H, x, y];
Hyy = D[H, y, y];
F1 = v;
```

**F1 is the velocity in the y direction.**
**F3 is the velocity in the x direction.**

```
F0 = (g + Hxx*u^2 + 2*Hxy*u*v + Hyy*v^2)/(1 + Hx^2 + Hy^2);
F2 = (-F0)*Hy - c*v;
F3 = u;
F4 = (-F0)*Hx - c*u;
p = Pi/5;
q = Pi/2;
a = 0;
B = 0.5;
c = 0;
g = 9.80665;
reps = {y -> y[t], v -> v[t], u -> u[t], x -> x[t], Derivative[1][x] -> Derivative[1][x][t],
  Derivative[1][y] -> Derivative[1][y][t]};
fraction = 0.9;
w = Hx*u + Hy*v;
totalenergy = 0.5*(u^2 + v^2 + w^2) + g*(B + H);
solv = Quiet[Solve[totalenergy - fraction*2*B*g == 0, v]];
```

**Note that the maximum energy of the board is 2 B g (setting mass equal to one).**
 **Any larger energy could allow the board to leave the surface which clearly is not part of this model.**
  **Thus fraction*2*B*g is the maximum energy for this experiment and Lorenz sets fraction equal to 0.9.**

```
TableForm[{Table[v /. solv /. {x -> 0, y -> y, u -> 4}, {y, 0.3, 0.36, 0.01}],
  {0.3, 0.31, 0.32, 0.33, 0.34, 0.35, 0.36}}]
```

| 0.719261 | 0.671687 | 0.619662 | 0.562012 | 0.496848 | 0.420771 | 0.32632 |
| -0.719261 | -0.671687 | -0.619662 | -0.562012 | -0.496848 | -0.420771 | -0.3263 |
| 0.3 | 0.31 | 0.32 | 0.33 | 0.34 | 0.35 | 0.36 |

```
sol3 = NDSolve[{Derivative[1][y][t] == (F1 /. reps), Derivative[1][x][t] == (F3 /. reps),
   Derivative[1][u][t] == (F4 /. reps), Derivative[1][v][t] == (F2 /. reps), y[0] == 0.3,
   u[0] == 4, v[0] == 0.7192609668778479, x[0] == 0}, {x[t], y[t], u[t], v[t]}, {t, 0, 5000},
  MaxSteps -> 15 000 000];

sol33 = NDSolve[{Derivative[1][y][t] == (F1 /. reps), Derivative[1][x][t] ==
    (F3 /. reps), Derivative[1][u][t] == (F4 /. reps), Derivative[1][v][t] == (F2 /. reps),
   y[0] == 0.33, u[0] == 4, v[0] == 0.5620115043136081, x[0] == 0}, {x[t], y[t], u[t], v[t]},
  {t, 0, 5000}, MaxSteps -> 15 000 000];

sol36 = NDSolve[{Derivative[1][y][t] == (F1 /. reps), Derivative[1][x][t] ==
    (F3 /. reps), Derivative[1][u][t] == (F4 /. reps), Derivative[1][v][t] == (F2 /. reps),
   y[0] == 0.36, u[0] == 4, v[0] == 0.32632627889791405, x[0] == 0}, {x[t], y[t], u[t], v[t]},
  {t, 0, 5000}, MaxSteps -> 15 000 000];

data3 = Partition[Flatten[Table[{x[t] /. sol3, y[t] /. sol3,
    (-a)*x[t] - B*Cos[p*x[t]]*Cos[q*y[t]]} /. sol3, {t, 0, 5000, 0.05}]], 3];
data3a = Table[{data3[[n]][[1]], data3[[n]][[2]], data3[[n]][[3]]}, {n, 1, 5000}];
plot3a = ListPointPlot3D[data3a, PlotStyle -> Directive[Red, PointSize[0.01]]];
```

```
data33 = Partition[Flatten[Table[{x[t] /. sol33, y[t] /. sol33,
    (-a)*x[t] - B*Cos[p*x[t]]*Cos[q*y[t]] /. sol33}, {t, 0, 5000, 0.05}]], 3];
data33a = Table[{data33[[n]][[1]], data33[[n]][[2]], data33[[n]][[3]]}, {n, 1, 5000}];
plot33a = ListPointPlot3D[data33a, PlotStyle -> Directive[Black, PointSize[0.01]]];

data36 = Partition[Flatten[Table[{x[t] /. sol36, y[t] /. sol36,
    (-a)*x[t] - B*Cos[p*x[t]]*Cos[q*y[t]] /. sol36}, {t, 0, 5000, 0.05}]], 3];
data36a = Table[{data36[[n]][[1]], data36[[n]][[2]], data36[[n]][[3]]}, {n, 1, 5000}];
plot36a = ListPointPlot3D[data36a, PlotStyle -> Directive[Blue, PointSize[0.01]]];
```

**The 3D views below show the similarity of the initial paths; the black path is confused by the reversal of the path at later times.**

```
threeDmogulview1 = Plot3D[1.02*H, {x, 0, 30}, {y, -10, 10}, ViewPoint -> {4., -0.95, 6.45},
  BoxRatios -> {1, 1, 0.05}, MeshStyle -> Opacity[0.3]];
Show[threeDmogulview1, plot3a, plot33a, plot36a]
```

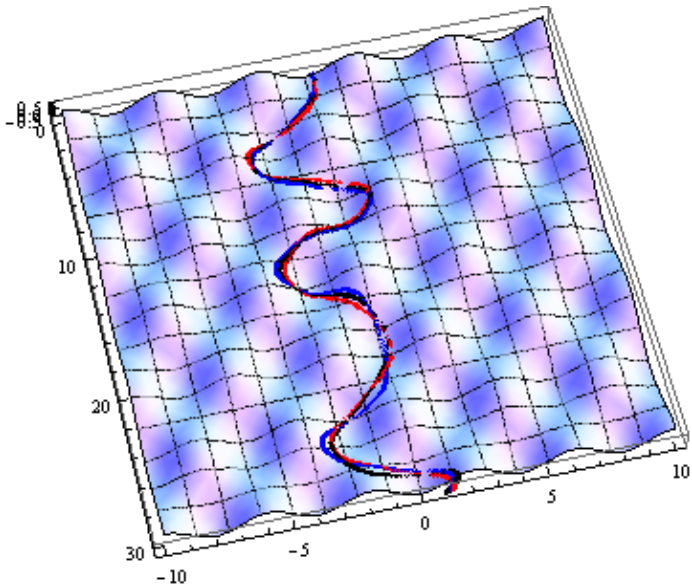

**A longer time plot clearly shows the erratic nature of the paths with no friction and no slope to the "hill".**

```
threeDmogulview2 = Plot3D[1.02*H, {x, 0, 220}, {y, -20, 30}, PlotPoints -> 50,
   ViewPoint -> {4., -0.95, 6.45}, BoxRatios -> {3, 1, 0.05}, Mesh -> None,
   PlotStyle -> Opacity[0.3]];
Show[threeDmogulview2, plot3a, plot33a, plot36a]
```

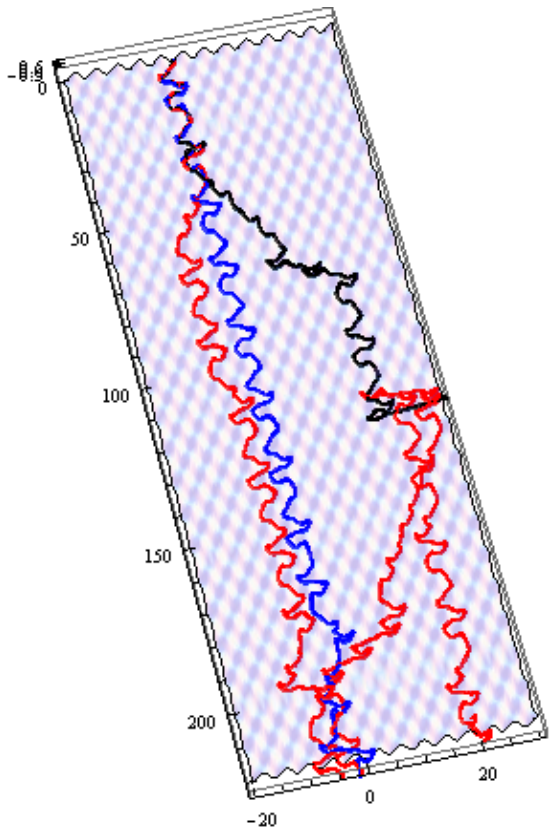

**The following generates Figure 20 :**

```
TableForm[
 {Table[v /. solv /. {x -> 0, y -> y, u -> 4}, {y, 0.34, 0.38, 0.02}], {0.34, 0.36, 0.38}}]
```

| | | |
|---|---|---|
| 0.496848 | 0.326326 | 0. + 0.191437 i |
| -0.496848 | -0.326326 | 0. - 0.191437 i |
| 0.34 | 0.36 | 0.38 |

```
sol = NDSolve[{Derivative[1][y][t] == (F1 /. reps), Derivative[1][x][t] == (F3 /. reps),
   Derivative[1][u][t] == (F4 /. reps), Derivative[1][v][t] == (F2 /. reps), y[0] == 0.36,
   u[0] == 4, v[0] == 0.32632627889791405, x[0] == 0},
    {x[t], y[t], u[t], v[t]}, {t, 0, 10000},
   MaxSteps -> 15000000];
table36 =
  Partition[Flatten[Table[{x[t] /. sol, y[t] /. sol, v[t] /. sol}, {t, 0, 10000, 0.01}]],
   3];
table36a =
  Table[{Mod[table36[[n]][[1]], 10], Mod[table36[[n]][[2]] - 2, 4] - 2, table36[[n]][[3]]},
   {n, 1, 1000001}];

table36a02a = Select[table36a, #1[[1]] < 0.02 &];
table36a98a = Select[table36a, #1[[1]] > 9.98 &];
plot36a02a = ListPlot[Table[{table36a02a[[n]][[2]], table36a02a[[n]][[3]]},
   {n, 1, Length[table36a02a]}],
    PlotStyle -> Directive[Red, PointSize[0.003]], PlotRange -> All,
   AspectRatio -> 1.2, Frame → True, Axes → False];
plot36a98a = ListPlot[Table[{table36a98a[[n]][[2]], table36a98a[[n]][[3]]},
   {n, 1, Length[table36a98a]}],
    PlotStyle -> Directive[Red, PointSize[0.003]], PlotRange -> All,
   AspectRatio -> 1.2, Frame → True, Axes → False];
Show[plot36a02a, plot36a98a]
```

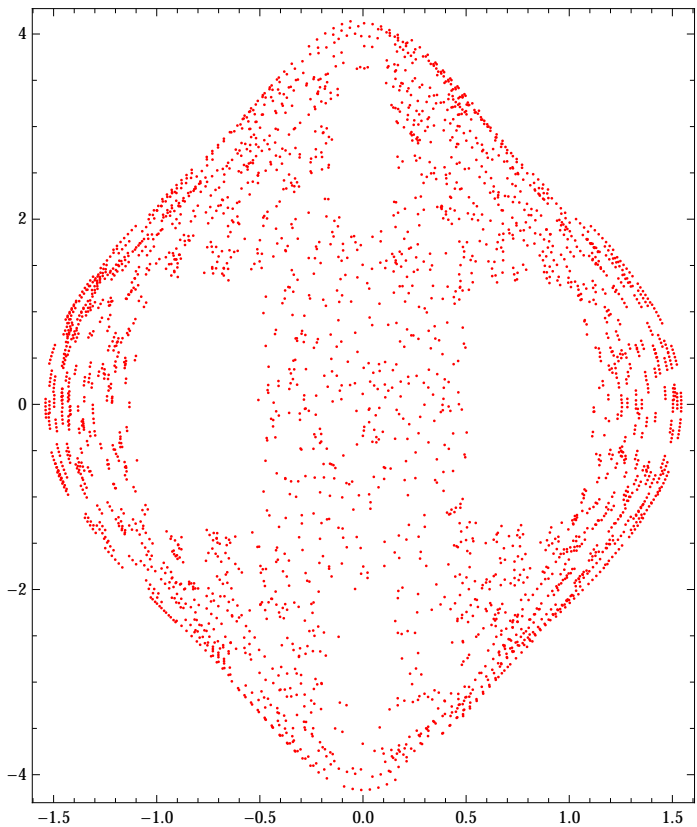

**For Figure 21, use starting y and v in one of the empty spaces:**

```
TableForm[{Table[v /. solv /. {x -> 0, y -> y, u -> 3}, {y, -0.8, -1, -0.1}], {-0.8, -0.9, -1}}]
```

| | | |
|---|---|---|
| 1.09726 | 0.4867 | 0. + 0.845076 𝕚 |
| -1.09726 | -0.4867 | 0. - 0.845076 𝕚 |
| -0.8 | -0.9 | -1 |

```
sol = NDSolve[{Derivative[1][y][t] == (F1 /. reps), Derivative[1][x][t] == (F3 /. reps),
   Derivative[1][u][t] == (F4 /. reps), Derivative[1][v][t] == (F2 /. reps), y[0] == -0.9,
   u[0] == 3, v[0] == 0.4866995724602631, x[0] == 0}, {x[t], y[t], u[t], v[t]}, {t, 0, 10000},
  MaxSteps -> 15000000];
table9 =
  Partition[Flatten[Table[{x[t] /. sol, y[t] /. sol, v[t] /. sol}, {t, 0, 10000, 0.01}]], 3];
table9a = Table[{Mod[table9[[n]][[1]], 10],
     Mod[table9[[n]][[2]] - 2, 4] - 2, table9[[n]][[3]]},
  {n, 1, 1000001}];
table9a02 = Select[table9a, #1[[1]] < 0.02 &];
table9a98 = Select[table9a, #1[[1]] > 9.98 &];
plot9a02a =
  ListPlot[Table[{table9a02[[n]][[2]], table9a02[[n]][[3]]}, {n, 1, Length[table9a02]}],
   PlotStyle -> Directive[Red, PointSize[0.003]], PlotRange -> {{-1.5, 1.5}, {-4, 4}},
   AspectRatio -> 1.2];
plot9a98 =
  ListPlot[Table[{table9a98[[n]][[2]], table9a98[[n]][[3]]}, {n, 1, Length[table9a98]}],
   PlotStyle -> Directive[Red, PointSize[0.003]], PlotRange -> {{-1.5, 1.5}, {-4, 4}},
   AspectRatio -> 1.2];
plotFigure21 = Show[plot9a02a, plot9a98]
```

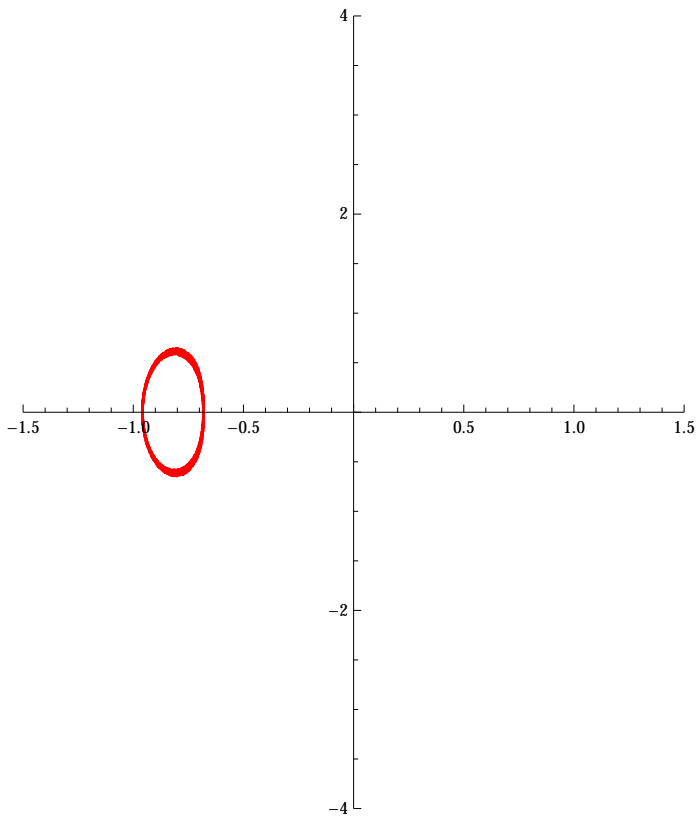

# 7. CONCLUSION

There are other very interesting developments of this board/sled model in the text that may be programmed by the reader. The book also contains the text of a previously unpublished talk given on December 29, 1972, entitled "Predictability: Does the Flap of a Butterfly's Wings in Brazil Set off a Tornado in Texas?" as well as chapters on "Our Chaotic Weather", "Encounters with Chaos" and "What Else is Chaos". These are all very worthwhile. There are brief descriptions of the Butterfly; the restricted three-body problem (which Lorenz refers to as Hill's reduced equations and is wonderfully covered in "Mathematica for Scientists and Engineers" by Richard Gass (ISBN 0132276127) and is shown at http://demonstrations.wolfram.com/RestrictedThreeBodyProblem/), and the Logistic Equation (covered in detail in many references, especially "Mathematical Navigator" by Heikki Ruskeepaa (ISBN 012603642X) and http://demonstrations.wolfram.com/ClassicLogisticMap/ and http://demonstrations.wolfram.com/DiscreteLogisticEquation/).

The Essence of Chaos is really a classic by a most innovative and original scientist and is available in paperback and Kindle from Amazon.com.
Dr. Edward N. Lorenz passed away on April 16, 2008.

# 8. About the author

Robert M.Lurie received a BS and ScD in Chemical Engineering at MIT (1952 and 1955) After retirement from a career in material development, he joined the Harvard Institute for Learning in Retirement where he has led or co-led seminars on Fractals and Chaos, Nonlinear Dynamics, The Essence of Chaos, Networks and Six Degrees of Separation, and Risk and Rationality. Mathematica was used for demonstrations in all of these seminars. Dr. Lurie was raised in central Maine, which is definitely a "snowy clime" where one may find mogul covered hills.

Address : 4 Tufts Road, Lexington, MA 02421
Email : RMLURIE@ALUM.MIT.EDU